\documentstyle[12pt]{article}

\begin{document}

\begin{center}
Continuum Field Model of Street Canyon: Theoretical Description
\end{center}
\begin{center}
Part 1
\end{center}
\begin{center}
Maciej M. Duras $\star$
\end{center}
\begin{center}
Institute of Physics, Cracow University of Technology, 
ulica Podchor\c{a}\.zych 1, PL-30-084 Cracow, Poland
\end{center}
\begin{center}
$\star$ Email: mduras @ riad.usk.pk.edu.pl
\end{center}

\begin{center}
{\em Engng. Trans.}  {\bf 48}, 415-432 (2000). 
\end{center}

\begin{abstract}
A general proecological urban road traffic control idea
for the street canyon is proposed
with emphasis placed on development of advanced
continuum field gasdynamical (hydrodynamical) 
control model of the street canyon.  
The continuum field model of optimal control of street canyon is studied.
The mathematical physics' approach (Eulerian approach) 
to vehicular movement,
to pollutants' emission, and to pollutants' dynamics is used.
The rigorous mathematical model is presented,
using gasdynamical (hydrodynamical) theory for both air constituents
and vehicles, including many types of vehicles and many types
of pollutant (exhaust gases) emitted from vehicles.
The six optimal control problems are formulated.
\end{abstract}

\section{Description of the model.}
\label{sect-description}
\setcounter{equation}{0}  
In the present article we develop a continuum field model of the street canyon.
In the next article we will deal with numerical examples 
\cite{Duras 1999 engtrans numer}.
The vehicular flow in the canyon is multilane bidirectional one-level
rectilinear, and it is considered
with two coordinated signalized junctions
\cite{Duras 1998 thesis, Duras 1997 PJES, Duras 1999 PJES}.
The vehicles belong to different vehicular classes:
passenger cars, and trucks.
Emissions from the vehicles are based on technical measurements
and many types of pollutants are considered
(carbon monoxide CO, hydrocarbons HC, nitrogen oxides ${\rm NO_{x}}$).
The vehicular dynamics is based on a hydrodynamical approach
\cite{Michalopoulos 1984}.
The governing equations are the continuity equation
for the number of vehicles, 
and Greenshields' equilibrium speed-density u-k model
\cite{Greenshields 1934}.

The model of dynamics of pollutants is also hydrodynamical.
The model consists of a set of mutually interconnected
nonlinear, three-dimensional, time-dependent, 
partial differential equations with nonzero right-hand sides (sources),
and of boundary and of initial problem.
The pollutants, oxygen, and the remaining gaseous constituents of air,
are treated as mixture of noninteracting, Newtonian, viscous
fluid (perfect or ideal gases).
The complete model incorporates as variables the following fields:
density of the mixture, mass concentrations of constituents of the mixture,
velocity of mixture, temperature of mixture, pressure of mixture,
intrinsic (internal) energy of mixture,
densities of vehicles, and velocities of vehicles.
The model is based on the assumption 
of local laws of balance (conservation) 
of: mass of the mixture, masses of its constituents,
momentum and energy of the mixture,
the numbers of the vehicles,
as well as of the state equations (Clapeyron's law and Greenshields' model).
The equations of dynamics are solved by the finite difference
scheme.

The six separate monocriterial optimization problems are formulated
by defining the functionals of total travel time,
of global emissions of pollutants, 
and of global concentrations of pollutants,
both in the studied street canyon
and in its two nearest neighbour substitute canyons.
The vector of control is a five-tuple
composed of two cycle times, two green times,
and one offset time between the traffic lights.
The optimal control problem consists of
minimization of the six functionals
over the admissible control domain.

\section{Assumptions.}
\label{sect-assumptions}
\setcounter{equation}{0}  
The {\bf geometrical assumptions} of the model are as follows:

{\bf G1.} The street canyon is represented by a cuboid:  
\begin{equation}
\Omega=[0, a] \times [0, b] \times [0, c],
\label{cuboid-def}   
\end{equation}
with the boundary $\partial \Omega$ composed of six walls:
\begin{equation}
\partial \Omega = \bigcup_{J={\rm I}}^{{\rm VI}} \partial \Omega_{J}.
\label{cuboid-border} 
\end{equation}
The structure of the canyon is simplified by the assumption 
that the walls of the buildings and the road surface are rectangles. 
If we put the origin of the Cartesian coordinate system $(x, y, z)$
in cuboid's corner, then we have 
two canyon walls $\Omega_{{\rm IV}}, \Omega_{{\rm V}},$ at $y=0, y=b$, 
and road's surface $\Omega_{{\rm VI}},$ at $z=0$.   
The remaining three non-solid open surfaces  
of air $\Omega_{{\rm I}}, \Omega_{{\rm II}}, \Omega_{{\rm III}},$ 
have the coordinates $x=0, x=a, z=c$, respectively.
We assume that always: 
\begin{equation}
(x, y, z) \in \Omega,
\label{cartesian-cuboid-point}   
\end{equation}
and
\begin{eqnarray}
& & \partial \Omega_{{\rm I}} = \{ (x, y, z) \in \Omega: x=0 \},
\label{cuboid-walls} \\
& & \partial \Omega_{{\rm II}} = \{ (x, y, z) \in \Omega: x=a \},
\nonumber \\
& & \partial \Omega_{{\rm III}} = \{ (x, y, z) \in \Omega: z=c \},
\nonumber \\
& & \partial \Omega_{{\rm IV}} = \{ (x, y, z) \in \Omega: y=0 \},
\nonumber \\
& & \partial \Omega_{{\rm V}} = \{ (x, y, z) \in \Omega: y=b \},
\nonumber \\
& & \partial \Omega_{{\rm VI}} = \{ (x, y, z) \in \Omega: z=0 \}.
\nonumber 
\end{eqnarray}

{\bf G2.} There are neither holes in the walls 
nor vegetation alongside the road. 
The remaining three surfaces of the cuboid also 
do not have holes since they simulate non-solid open rectangles of air. 

{\bf G3.} The road sections which constitute 
the bottom of the street canyon are rectilinear. 

{\bf G4.} At each end of the street canyon there are entrance 
and exit junctions ($M = 2$) with traffic signals 
(their coordinates are $x=0, x=a$). 

{\bf G5.}  The vehicles of $VT$ distinguishable emission types  
are material points.
The vehicles are treated as hydrodynamical fluid. 
There are $n_{{\rm L}}=n_{1}$ left lanes and $n_{2}=n_{{\rm R}}$ right ones 
(the traffic is bidirectional).

The {\bf physical assumptions} of this model are as follows
(cf. papers \cite{Eringen 1990} - \cite{Saltanov 1984}):

{\bf P1.} The  considered mixture of gases consists of 
$N=N_{E}-1 +N_{A}$  gases. 
The first $N_{E}-1=3$ gases are the exhaust gases emitted by  
vehicle engines during combustion 
(${\rm CO, CH, NO_{x}}$, we neglect the presence of ${\rm SO_{2}}$). 
The remaining $N_{A}=9$ gases are the constituents of air: 
${\rm O_{2}, N_{2}, Ar, CO_{2}, Ne, He, Kr, Xe, H_{2}}$
(we neglect the presence of ${\rm H_{2}O, O_{3}}$). 

{\bf P2.} The walls of the canyon and the surface of the road 
are impervious for all gases of the mixture. 
The remaining three surfaces of the cuboid 
are pervious for external fluxes of exhaust gases and air constituents. 

{\bf P3.} The internal sources of air constituents 
are not present with the exception of oxygen, {\it i.e.}, 
$N_{E}$ constituent of the gaseous mixture. 
There are internal mobile sources of exhaust 
gases (passenger cars and trucks, with many types of engines: 
diesel or petrol, and with different ages of engines). 
During combustion, the engine consumes oxygen, therefore with each 
internal mobile source of exhaust gases, 
a negative source of oxygen (sink) is connected.

{\bf P4.} The gaseous mixture is treated as a compressible, Newtonian, 
and viscous fluid. We assume that also the constituents 
of the mixture are  compressible, Newtonian, and viscous fluids. 
The constituents do not interact with each other. 
The $i$th  constituent possesses individual 
velocity ${\bf{v}}_{i}(x, y, z, t)$, density $\rho_{i}(x, y, z, t)$, 
and pressure $p_{i}(x, y, z, t)$, whereas the mixture 
possesses total velocity
${\bf{v}}(x, y, z, t)$, density$\rho(x, y, z, t)$, and pressure 
$p(x, y, z, t)$.
We assume that
\begin{eqnarray}
& & {\bf{v}}(x, y, z, t)=
\sum_{i=1}^{N}\frac{\rho_{i}(x, y, z, t)}{\rho(x, y, z, t)}
\cdot {\bf{v}}_{i}(x, y, z, t), 
\label{v-rho-p-sums-1} \\
& & \rho(x, y, z, t)=
\sum_{i=1}^{N}\rho_{i}(x, y, z, t), 
\label{v-rho-p-sums-2} \\
& & p(x, y, z, t)=
\sum_{i=1}^{N}p_{i}(x, y, z, t). 
\label{v-rho-p-sums-3}
\end{eqnarray}
In order to simplify the set of equations governing 
the dynamics of mixture,
we assume that the total velocity 
is equal to the velocities of the constituents 
\begin{equation}
{\bf{v}}(x, y, z, t)=
{\bf{v}}_{i}(x, y, z, t), i=1, ..., N.
\label{vel-equal}
\end{equation}
 
Hence, we can restrict our attention to the equations 
of balance of total momentum of mixture {\bf E1}, 
of the total mass of mixture {\bf E2}, 
of masses of constituents {\bf E3}, 
and of the energy of mixture {\bf E4}. 
We assume that the equation of state for mixture {\bf E5}  
is averaged over the constituents. 
We also consider equations of balances of the numbers of vehicles {\bf E6},
as well as equations of state for vehicles {\bf E7}. 

\section{Variables.}
\label{sect-variables}
\setcounter{equation}{0}  
The following set of descriptive dynamic model 
variables {\bf A0-A10} together with their 
boundary {\bf B0-B8} (for $t \geq 0$),
and initial conditions {\bf C0-C7} (for $t=0$), 
and with the set of equations {\bf E1-E8} 
that governs their dynamics, is assumed \cite{Duras 1998 thesis}.
We always consider 
\begin{equation}
(x, y, z, t) \in \Sigma, \Sigma= \Omega \times [0, T_{S}],
\label{manifold-omega-sigma}
\end{equation}
where $\Sigma$ is manifold of domains of the fields, $T_{S} > 0$ is time of simulation.
The border $\partial \Sigma$ of $\Sigma$
is composed of six subsets $\partial \Sigma_{J}$:
\begin{eqnarray}
& & \partial \Sigma = \bigcup_{J={\rm I}}^{{\rm VI}} \partial \Sigma_{J},
\label{sigma-border} \\ 
& & \partial \Sigma_{J} = \partial \Omega_{J} \times [0, T_{S}],
J= {\rm I, ..., VI}.
\nonumber 
\end{eqnarray}

{\bf A0.} $T(x, y, z, t)$, temperature of the gaseous mixture.
 
{\bf A1.} ${\bf{v}}(x, y, z, t)$, total velocity of the gaseous mixture
(compare Eq. (\ref{v-rho-p-sums-1})). 
 
{\bf A2.} $\rho(x, y, z, t)$, total density of the gaseous mixture
(compare Eq. (\ref{v-rho-p-sums-2})). 
  
{\bf A3.} $c_{i}(x, y, z, t)$, mass concentration of 
the $i$th constituent of gaseous mixture, 
\begin{equation}
c_{i}(x, y, z, t)=\frac{\rho_{i}(x, y, z, t)}{\rho(x, y, z, t)},
i=1, ..., N
\label{conc-def}
\end{equation}   
(let us note that due to the condition  
\begin{equation}
\sum_{i=1}^{N}c_{i}(x, y, z, t)=1,
\label{conc-sum}
\end{equation}
one concentration of the constituent is a dependent variable). 

{\bf A4a.} $p(x, y, z, t)$, pressure of gaseous mixture
(compare Eqs (\ref{v-rho-p-sums-3}), (\ref{state-eq})).
 
{\bf A4b.} $p_{i}(x, y, z, t), i=1, ..., N,$ 
partial pressure of the $i$th constituent of gaseous mixture
(compare Eqs. (\ref{v-rho-p-sums-3}), 
(\ref{state-eq}), (\ref{constituent-state-eq})).

{\bf A5.} $k_{l, vt}^{s}(x, t),$
density of vehicles of type $vt$ on the $l$th lane 
measured in $[\frac{{\rm veh}}{{\rm m}}]$, 
where for $n_{1}=n_{L}$ lanes $s=1$ and $l$ is the left lane's number, 
$l=1, ..., n_{L},$
whereas 
for $n_{2}=n_{R}$ lanes $s=2$ and $l$ is the right lane's number,
$l=1, ..., n_{R},$
$vt$ is the vehicular type number, $vt=1, ..., VT$.

{\bf A6.} ${\bf{w}}_{l, vt}^{s}(x, t),$    
velocity of vehicles of type $vt$ on the $l$th lane.

{\bf A7.} $e_{l, ct, vt}^{s}(x, t),$   
emissivity of $ct$th constituent of exhaust gases 
from vehicles of type $vt$ on $l$th lane measured in 
$[\frac{{\rm kg}}{{\rm m \cdot s}}]$,
$ct$ is number of constituent, $ct=1, ..., CT$.
 
{\bf A8.} ${\bf{u}}_{m}=(g_{m}, C_{m}, F),$ 
vector of control on the $m$th  junction, $m=1, ..., M$ ($M=2$), 
which contains traffic signals  
green times $g_{m}$,
and cycle times $C_{m}$,
and offset time $F$ between the traffic signals.
The vector of control ${\bf{u}}$ reads:
\begin{equation}
{\bf {u}}=(g_{1}, C_{1}, g_{2}, C_{2}, F).
\label{5-tuple-def}
\end{equation} 
The admissible control domain set $U^{{\rm adm}}$ 
for this vector in the simulation 
time period $T_{S}$ reads:
\begin{eqnarray}
& & U^{{\rm adm}}=\{ (g_{1}, C_{1}, g_{2}, C_{2}, F): 
\label{U-admissible-def}\\
& & g_{m} \in (g_{m, {\rm min}}, g_{m, {\rm max}}),
C_{m} \in (C_{m, {\rm min}}, C_{m, {\rm max}}),
F \in (F_{{\rm min}}, F_{{\rm max}}), m=1, ..., M \}, 
\nonumber
\end{eqnarray}
whereas 
$g_{m, {\rm max}}= C_{m}-g_{m, {\rm orth}}$,
where $g_{m, {\rm orth}}$ are green times on orthogonal direction of the junctions
(on the canyons orthogonal to the one studied),
$F_{{\rm max}}= C_{2} - \delta_{F}$,
where $\delta_{F}$ is unit step in direction of $F$ in parameter space 
(compare {\bf F0}).

{\bf A9.} $G_{{\rm out}}^{m}(g_{m}, C_{m}, F, t)$, traffic signal on $m$th junction 
at $x=x_{m}$.
For $x_{1}= a$ the traffic signal $G_{{\rm out}}^{1}$ governs all left lanes 
(outgoing vehicles) and all right lanes (incoming vehicles),
whereas for $x_{2}= 0$ the traffic signal $G_{{\rm out}}^{2}$ governs
all right lanes (outgoing vehicles) and all left lanes (incoming vehicles). 
For the signals we assume the Boolean output values:
${\rm GREEN}$ and ${\rm RED}$.

{\bf A10.} $\sigma_{l, vt}^{s}(x, t),$  
rate of change of linear density of energy connected to heat produced 
by engines of vehicles of type $vt$ on $l$th lane measured 
in $[\frac{{\rm J}}{{\rm m \cdot s}}]$.
 
\section{Boundary conditions.}
\label{sect-boundary}
\setcounter{equation}{0}  
We assume the following boundary conditions \cite{Duras 1998 thesis}:

{\bf B0a-B0c.} $T|_{\partial \Sigma_{K}} = T_{K}|_{\Sigma_{K}^{'}}$.  

{\bf B0d-B0f.} $\nabla_{{\bf{n}}} T|_{\partial \Sigma_{L}} = \vec{0}$.

{\bf B1a-B1c.} ${\bf{v}}|_{\partial \Sigma_{K}} = {\bf{v}}_{K}|_{\Sigma_{K}^{'}}$. 

{\bf B1d-B1f.} $\nabla_{{\bf{n}}} {\bf{v}}|_{\partial \Sigma_{L}} = {\bf O}$.

{\bf B2a-B2c.} $\rho|_{\partial \Sigma_{K}} = \rho_{K}|_{\Sigma_{K}^{'}}$.  

{\bf B2d-B2f.} $\nabla_{{\bf{n}}} \rho|_{\partial \Sigma_{L}} = \vec{0}$.

{\bf B3a-B3c.} $c_{i}|_{\partial \Sigma_{K}} = c_{i, K}|_{\Sigma_{K}^{'}}$.

{\bf B3d-B3f.} $\nabla_{{\bf{n}}} c_{i}|_{\partial \Sigma_{L}} = \vec{0}$.

{\bf B4a-B4c.} $p|_{\partial \Sigma_{K}} = p_{K}|_{\Sigma_{K}^{'}}$. 

{\bf B4d-B4f.} $\nabla_{{\bf{n}}} p|_{\partial \Sigma_{L}} = \vec{0}$.

{\bf B5a-B5d.} $k_{l, vt}^{s}(x_{m}, t)=k_{l, vt, P}^{s}(t)$.

{\bf B6a-B6d.} ${\bf{w}}_{l, vt}^{s}(x_{m}, t)={\bf{w}}_{l, vt, P}^{s}(t)$.

{\bf B7a-B7d.} $e_{l, ct, vt}^{s}(x_{m}, t)=e_{l, ct, vt, P}^{s}(t)$.

{\bf B8a-B8d.} $\sigma_{l, vt}^{s}(x_{m}, t)=\sigma_{l, vt, P}^{s}(t)$.

We define additional sets:
\begin{eqnarray}
& & \Sigma_{{\rm I}}^{'} = \{ (y, z, t): (x, y, z, t) \in \Sigma \}
= \Sigma_{{\rm II}}^{'},
\label{sigma-border-special} \\
& & \Sigma_{{\rm III}}^{'} = \{ (x, y, t): (x, y, z, t) \in \Sigma \},
\nonumber 
\end{eqnarray}
and we assume that $K={\rm I, II, III},$ $L={\rm IV, V, VI}$.
The gradient operator $\nabla_{\bf{n}}$ works
in direction of unit normal outward vector $\bf{n}$ 
to border $\partial \Sigma_{L}$.
$\nabla_{\bf{n}} \bf{v}$ is gradient of vector (so it is a tensor of rank 2), 
${\bf O}$ is a zero tensor.
$P={\rm in, out}$, is the input and output index, and we have the following
combinations of triads of indices: 
$(s, P, m)=(1, {\rm in}, 2), (2, {\rm in}, 1), (1, {\rm out}, 1),(2, {\rm out}, 2)$,
respectively for {\bf B5-B8}.
Conditions {\bf B1d-B1f} result from viscosity of the gaseous mixture 
since the velocity of viscous 
fluid on immobile and impervious surface is zero. 
Similarly, conditions {\bf B0d-B0f}, {\bf B2d-B2f}, 
{\bf B3d-B3f}, {\bf B4d-B4f} result from the fact 
that the walls and the surface of the road are impervious solid bodies.
According to \cite{Michalopoulos 1984},
we assume the boundary conditions {\bf B5a-B5d}
in the form: 

{\bf B5aS-B5dS.} 

$k_{l, vt, P}^{s}(t)=k_{l, vt, {\rm arrival}}^{s},$ 
if $G_{{\rm out}}^{m}(C_{m}, g_{m}, F, t)={\rm GREEN},$ 
and ${\rm QUEUE}(x_{m})={\rm FALSE}$,

$k_{l, vt, P}^{s}(t)=k_{l, vt, {\rm sat}}^{s},$ 
if $G_{{\rm out}}^{m}(C_{m}, g_{m}, F, t)={\rm GREEN},$ 
and ${\rm QUEUE}(x_{m})={\rm TRUE}$,

$k_{l, vt, P}^{s}(t)=k_{l, vt, {\rm jam}}^{s},$ 
if $G_{{\rm out}}^{m}(C_{m}, g_{m}, F, t)={\rm RED}$,

where ${\rm QUEUE}(x_{m})={\rm TRUE/FALSE}$ means
that there exist/does not exist a queue at $x=x_{m}$, 
and $k_{l, vt, {\rm arrival}}^{s}$,
$k_{l, vt, {\rm sat}}^{s}$,
and $k_{l, vt, {\rm jam}}^{s}$,
are arrival, saturation, and jam vehicular densities, respectively
(compare Tables 3, and 8 of \cite{Duras 1999 engtrans numer}).

The existence of the queues at the entrances to the canyon 
(at $x_{2}=0$ for the left lanes, 
and  at $x_{1}=a$ for the right lanes) 
is determined by the values of the vehicular densities changing in 
the following way:

{\bf B5aSS-B5dSS.}
\begin{eqnarray}
& & k_{l, vt}^{s}(\xi_{s}, t)=k_{l, vt, Q}^{s},
{\rm for \,} t \in A_{Q}^{s},
\label{k-kGREEN-0} \\
& & A_{{\rm GREEN}}^{s} =
[0, T_{S}] \cap
\bigcup_{n=-\infty}^{+\infty} [ t_{s} + n C_{1}, t_{s} + n C_{1}+g_{1}),  
\label{k-kGREEN-1} \\
& & A_{{\rm RED}}^{s} =
[0, T_{S}] - A_{{\rm GREEN}}^{s},
Q={\rm GREEN, RED},
\label{k-kGREEN-2} \\
& & \xi_{1}=-\delta_{x}, \xi_{2}=a+\delta_{x},
t_{1}=0, t_{2}=F,
\label{k-kGREEN-3} \
\end{eqnarray}
where $k_{l, vt, {\rm GREEN}}^{s}, k_{l, vt, {\rm RED}}^{s}, $
are green and red vehicular densities, respectively
\cite{Duras 1999 engtrans numer},
and $\delta_{x}$ is the unit step in x-direction in the domain space $\Sigma$.

Remark: The functions: 
$T_{J}, {\bf{v}}_{J}, \rho_{J}, c_{i, J}, p_{J},
k_{l, vt, P}^{s}, {\bf{w}}_{l, vt, P}^{s}, e_{l, ct, vt, P}^{s},  
\sigma_{l, vt, P}^{s}, $
are given and they fulfill 
the natural constraints:
\begin{equation}
\sum_{i=1}^{N}c_{i, J}(x, y, z, t)=1.
\label{conc-sum-boundary}
\end{equation}
  
\section{Initial conditions.}
\label{sect-initial}
\setcounter{equation}{0}  
The initial conditions are as follows \cite{Duras 1998 thesis}:

{\bf C0.} $T|_{\Sigma_{0}}=T_{0}|_{\Omega}.$

{\bf C1.} ${\bf{v}}|_{\Sigma_{0}}={\bf{v}}_{0}|_{\Omega}.$

{\bf C2.} $\rho|_{\Sigma_{0}}=\rho_{0}|_{\Omega}.$

{\bf C3.} $c_{i}|_{\Sigma_{0}}=c_{i, 0}|_{\Omega}.$

{\bf C4.} $p|_{\Sigma_{0}}=p_{0}|_{\Omega}.$

{\bf C5a-C5b.} $k_{l, vt}^{s}(x, 0)=k_{l, vt, 0}^{s}(x).$

{\bf C6a-C6b.} ${\bf{w}}_{l, vt}^{s}(x, 0)={\bf{w}}_{l, vt, 0}^{s}(x).$

{\bf C7a-C7b.} $e_{l, ct, vt}^{s}(x, 0)=e_{l, ct, vt, 0}^{s}(x).$

{\bf C8a-C8b.} $\sigma_{l, vt}^{s}(x, 0)=\sigma_{l, vt, 0}^{s}(x).$

Remark: The functions: 
$T_{0}, {\bf{v}}_{0}, \rho_{0}, c_{i, 0}, p_{0},
k_{l, vt, 0}^{s}, {\bf{w}}_{l, vt, 0}^{s}, e_{l, ct, vt, 0}^{s}, 
\sigma_{l, vt, 0}^{s},$
are given and they fulfill the natural constraint:
\begin{equation}
\sum_{i=1}^{N}c_{i, 0}(x, y, z, t)=1.
\label{conc-sum-initial}
\end{equation}
  
\section{Sources.}
\label{sect-sources}
\setcounter{equation}{0}  
In order to represent the emission process, 
we assume the following internal sources \cite{Duras 1998 thesis}:

{\bf D0.} $\sigma(x, y, z, t)$, the rate of change of the volume density 
of internal sources of energy connected with the production 
of heat by vehicular engines, measured in 
$[\frac{{\rm J}}{{\rm m \cdot s}}]$. 
We assume that the sources of energy are situated 
in $n_{s}$  left and right lanes at $y=y_{l}^{s},$  
at the level of the road $z=0$:
\begin{equation}
\sigma(x, y, z, t)=\frac{1}{b \cdot c} 
\cdot
\sum_{s=1}^{2}
\sum_{l=1}^{n_{s}}\sum_{vt=1}^{VT}\sigma_{l, vt}^{s}(x, t)
\chi_{D_{l}^{s}}(x, y, z),
\label{sigma-source-def}
\end{equation} 
where
\begin{equation}
D_{l}^{s}= \{ (x, y, 0): (x, y, 0) \in \Omega, y=y_{l}^{s} \},
\label{Dl-def} 
\end{equation}
are the vehicular lanes, and
\begin{equation}
\chi_{D}(x, y, z)=
\left\{
  \begin{array}{ll}
   1 & \mbox{for $(x, y, z) \in D$} \\
   0 & 
   \mbox{for $(x, y, z) \notin D$}
  \end{array}
\right.
,
\label{chi-function-def}
\end{equation}
is the characteristic function of set $D$. 

{\bf D1.} $S(x, y, z, t)$, the rate of change of the volume 
density of internal sources of gaseous 
mixture consisting of exhaust gases and oxygen, measured in 
$[\frac{{\rm kg}}{{\rm m \cdot s}}]$.
 
{\bf D2.} $S^{E}_{ct}(x, y, z, t)$,  
the rate of change of the volume density of internal sources 
(the emission rate) of the $ct$th constituent of exhaust gases emitted 
by all vehicles in the canyon, measured in  
$[\frac{{\rm kg}}{{\rm m \cdot s}}]$.
We assume that the sources of exhaust gases are situated 
in $n_{s}$ left and right lanes $y=y_{l}^{s},$
at the level of the road $z=0$:
\begin{equation}
S^{E}_{ct}(x, y, z, t)=\frac{1}{b \cdot c} \cdot 
\sum_{s=1}^{2}
\sum_{l=1}^{n_{s}}\sum_{vt=1}^{VT}e_{l, ct, vt}^{s}(x, t)
\chi_{D_{l}^{s}}(x, y, z) .
\label{Set-source-def} 
\end{equation}  

$S^{E}_{N_{E}}(x, y, z, t)$,  
the volume density of negative internal sources (the emission rate) 
of oxygen absorbed by  all vehicles in the canyon, measured 
in $[\frac{{\rm kg}}{{\rm m \cdot s}}]$. We assume that  
\[
S^{E}_{N_{E}}(x, y, z, t)={\rm ONOX} \cdot S^{E}_{N_{E}-1}(x, y, z, t),
\]
where ${\rm ONOX}=-0.5308$.
The following relation holds:
\begin{equation}
S(x, y, z, t)=\sum_{ne=1}^{N_{E}}S^{E}_{ne}(x, y, z, t).
\label{sources-sum}
\end{equation}  

\section{Equations of dynamics.}
\label{sect-dynamics}
\setcounter{equation}{0}  
Under the above model specifications, 
the  complete set of equations of dynamics of the model 
is formulated as follows
(we follow the general idea presented  
in \cite{Eringen 1990, Landau 1986}):

{\bf E1. Balance of momentum of mixture - Navier Stokes equation.}
\begin{equation}
\rho (\frac{\partial {\bf{v}}}{\partial t}
+ ({\bf{v}} \circ \nabla) {\bf{v}}) + S{\bf{v}}=
- \nabla p + \eta \Delta {\bf{v}} 
+ (\xi + \frac{\eta}{3}) \nabla ({\rm div} {\bf{v}}) + {\bf{F}},
\label{Navier-Stokes-eq}
\end{equation}
where $\eta$ is the first viscosity coefficient 
($\eta=18.1 \cdot 10^{-6} [\frac{{\rm kg}}{{\rm m \cdot s}}]$ 
for air at temperature $T=293.16$ [K]), 
$\xi$ is the second viscosity coefficient 
($\xi=15.6 \cdot 10^{-6} [\frac{{\rm kg}}{{\rm m \cdot s}}]$ 
for air at temperature $T=293.16$[K], compare \cite{Prangsma 1973}), 
${\bf{F}}=\rho {\bf{g}}$ is the gravitational body force density,
${\bf{g}}$ is the gravitational acceleration 
of Earth (${\bf{g}}=(0, 0, -9.81) [\frac{{\rm m}}{{\rm s^{2}}}]$),
$\nabla \bf{v}$ is gradient of the vector (so it is a tensor of rank 2). 
We assume that the gaseous mixture is a compressible and 
viscous fluid.

{\bf E2. Balance of mass of mixture - Equation of continuity.}
\begin{equation}
\frac{\partial \rho}{\partial t}
+ {\rm div}(\rho {\bf{v}})=S.
\label{continuity-eq}
\end{equation}  

We have assumed the source {\bf D1}.

{\bf E3. Balances of masses of constituents of mixture - Diffusion equations.}
 
{\bf E3a.}
\begin{eqnarray}
& & \rho (\frac{\partial c_{i}}{\partial t} + {\bf{v}} \circ \nabla c_{i})=
S^{E}_{i} - c_{i} S + 
\label{diffusion-eq-a} \\
& & + \sum_{m=1}^{N-1}\{ (D_{im}-D_{iN})
\cdot {\rm div}[\rho \nabla (c_{m}+\frac{k_{T, m}}{T} \nabla T)] \},
i=1, ..., N_{E}. \nonumber
\end{eqnarray}

{\bf E3b.}
\begin{eqnarray}
& & \rho (\frac{\partial c_{i}}{\partial t} + {\bf{v}} \circ \nabla c_{i})=
- c_{i} S +
\label{diffusion-eq-b} \\
& & + \sum_{m=1}^{N-1}\{ (D_{im}-D_{iN})
\cdot {\rm div}[\rho \nabla (c_{m}+\frac{k_{T, m}}{T} \nabla T)] \},
i=(N_{E}+1), ..., N, \nonumber
\end{eqnarray}
where $D_{im}=D_{mi}$ is the mutual diffusivity coefficient 
from the $i$-th constituent to $m$-th one, and $D_{ii}$  is the 
autodiffusivity coefficient of the $i$-th constituent,
and $k_{T, m}$ is the thermodiffusion ratio of the $m$-th constituent. 
The diffusivity coefficients and thermodiffusion ratios 
are constant and known 
(compare \cite{Chapman 1970}).
In {\bf E3a} we have assumed the sources {\bf D1-D2}. 
In {\bf E3b} only the source {\bf D1} is taken into account. 
Since the mixture is in motion, we cannot neglect the convection term:
${\bf{v}} \circ \nabla c_{i}$.  
We assume that the barodiffusion and gravitodiffusion coefficients 
are equal to zero.

{\bf E4. Balance of energy of mixture.}
\begin{equation}
\rho (\frac{\partial \epsilon}{\partial t} + {\bf{v}} \circ \nabla \epsilon)=
-(-\frac{1}{2}{\bf{v}}^{2} + \epsilon) S
+ {\bf T}:\nabla {\bf{v}}+{\rm div}(-{\bf{q}})+\sigma,
\label{energy-eq} 
\end{equation}
where $\epsilon$ is the mass density of intrinsic (internal) energy 
of the air mixture, 
${\bf T}$ is the stress tensor, 
symbol $:$ denotes the contraction operation,
$\bf{q}$ is the vector of flux of heat.
We assume that \cite{Duras 1998 thesis}:
\begin{eqnarray} 
& & \epsilon=\sum_{i=1}^{N}\epsilon_{i},
\label{intrinsic-energy-def} \\
& & \epsilon_{i}=
\frac{1}{m_{i}} 
\{ 
c_{i} k_{B} T 
\exp(-\frac{m_{i}|{\bf{g}}|z}{k_{B}T}) \cdot
[
(-\frac{z}{c}) \cdot (1-\exp(-\frac{m_{i}|{\bf{g}}|c}{k_{B}T}))
-\exp(-\frac{m_{i}|{\bf{g}}|c}{k_{B}T})
]
\} +
\label{ith-intrinsic-energy-def} \\
& & + \tilde{\mu}_{i}c_{i}, 
\nonumber \\
& &
\tilde{\mu}_{i}=\frac{\mu_{i}}{m_{i}},
\label{chemical-potential-tilde} \\
& & 
\mu_{i}= 
k_{B} T \cdot
\{
\ln
[
(c_{i} p) (k_{B} T)^{-\frac{c_{p, i}}{k_{B}}}
(\frac{m_{{\rm air}}}{m_{i}}) (\frac{2\pi h^{2}}{m_{i}})^{\frac{3}{2}}
]
\}
+m_{i} |{\bf{g}}| z
,
\label{chemical-potential-def} \\
& & {\rm T}_{mk}=-p\delta_{mk}
+
\label{stress-tensor-def} \\
& & + \eta \cdot 
[
(
\frac{\partial v_{m}}{\partial x_{k}}
+
\frac{\partial v_{k}}{\partial x_{m}}
-
\frac{2}{3} \delta_{mk} {\rm div}({\bf{v}})
)
\frac{\partial v_{k}}{\partial x_{m}}
]
+
\xi \cdot
[
(
\delta_{mk} {\rm div}({\bf{v}})
)^{2}
],
m, k=1, ...,3,
\nonumber \\
& & {\bf T}:\nabla {\bf{v}}=
\sum_{m=1}^{3} \sum_{k=1}^{3} 
{\rm T}_{mk} \frac{\partial v_{m}}{\partial x_{k}}, 
\label{stress-tensor-velocity-contraction} \\
& & {\bf{q}}=
\sum_{i=1}^{N}
\{
[
(\frac{\beta_{i}T}{\alpha_{ii}}
+ \tilde{\mu}_{i}) {\bf{j}}_{i}
]
+
[
(-\kappa) \nabla T
]
\}
,
\label{vector-flux-heat-def} \\
& & {\bf{j}}_{i}=
-\rho D_{ii} (c_{i}+\frac{k_{T, i}}{T} \nabla T)
,
\label{vector-flux-mass-def} \\
& & \alpha_{ii}=
\frac{
[\rho D_{ii}]
}
{ 
[(\frac{\partial \tilde{\mu}_{i}}{\partial c_{i}})
_{(c_{n})_{n=1, ..., N, i \neq n}, T, p}]
},
\label{alpha-def} \\
& & \beta_{i}=
[\rho D_{ii}] \cdot
\{
\frac{k_{T, i}}{T}
-
\frac{
[(\frac{\partial \tilde{\mu}_{i}}{\partial T})
_{(c_{n})_{n=1, ..., N}, p}]
}
{
[(\frac{\partial \tilde{\mu}_{i}}{\partial c_{i}})
_{(c_{n})_{n=1, ..., N, i \neq n}, T, p}]
}
\}
,
\label{beta-def}
\end{eqnarray}
where 
$\epsilon_{i}$ is the mass density of intrinsic (internal) energy 
of the $i$th constituent of the air mixture,
$m_{i}$ the molecular mass of the $i$th constituent,
$k_{B}=1.3807 \cdot 10^{-23} [\frac{{\rm J}}{{\rm kg}}]$ 
is Boltzmann's constant,
$\mu_{i}$ is the complete partial chemical potential of the $i$th constituent 
of the air mixture (it is complete since it is composed of
chemical potential without external force 
field and of external potential),
$m_{{\rm air}}=28.966$ [u] is the molecular mass of air
($1[{\rm u}]=1.66054 \cdot 10^{-27}$ [kg]),
$\delta_{mk}$ is Kronecker's delta,
$c_{p, i}$ is the specific heat at constant pressure of the $i$th 
constituent of air mixture,
$h=6.62608 \cdot 10^{-34}$ [$J \cdot s$] is Planck's constant,
${\bf{j}}_{i}$ is the vector of flux of mass of the $i$th constituent
of the air mixture,
and $\kappa$ is the coefficient of thermal conductivity of air.
These magnitudes are derived from
Grand Canonical ensemble with external gravitational Newtonian field.

{\bf E5. Equation of state of the mixture - Constitutive equation -
Clapeyron's equation.}

\begin{equation}
\frac{p}{\rho}=\frac{R}{m_{{\rm air}}} \cdot T
\label{state-eq}
\end{equation}
is Clapeyron's equation of state for a gaseous mixture,
where $R=8.3145$ [$\frac{{\rm J}}{{\rm mole \cdot K}}$] is the gas constant.
\begin{equation}
p_{i}=c_{i} \cdot \frac{m_{{\rm air}}}{m_{i}} \cdot p
\label{constituent-state-eq}
\end{equation}
are partial pressures of constituents according to Dalton's law.

{\bf E6. Balances of numbers of vehicles - Equations of continuity of vehicles.}

\begin{equation}
\frac{\partial k_{l, vt}^{s}}{\partial t} 
+ {\rm div}(k_{l, vt}^{s}{\bf{w}}_{l, vt}^{s})=0.  
\label{vehicle-continuity-eq-a}
\end{equation}  

{\bf E7. Equations of state of vehicles - Greenshields model.}

\begin{equation}
{\bf{w}}_{l, vt}^{s}(x, t)=
(
w_{l, vt, f}^{s} \cdot
(1-\frac{k_{l, vt}^{s}(x, t)}{k_{l, vt, {\rm jam}}^{s}}), 0, 0).
\label{Greenshields-eq-a} 
\end{equation}

The Greenshields equilibrium speed-density u-k model is assumed 
\cite{Greenshields 1934}. 
The values of maximum free flow speed 
$w_{l, vt, f}^{s}$, 
and of jam vehicular densities 
$k_{l, vt, {\rm jam}}^{s}$,
are given in Tables 3 and 8 of \cite{Duras 1999 engtrans numer}.
 
{\bf E8. Technical parameters.}

The dependence of emissivity on the density and velocity 
of vehicles is assumed in the form \cite{Sturm 1996}:

{\bf E8a.}

\begin{equation}
e_{l, ct, vt}^{s}(x, t)=
k_{l, vt}^{s}(x, t) \cdot
[
(\frac{|{\bf{w}}_{l, vt}^{s}(x, t)|-\tilde{w}_{ct, vt, i_{l}}}
{\tilde{w}_{ct, vt, i_{l}+1}-\tilde{w}_{ct, vt, i_{l}}}) \cdot
(\tilde{e}_{ct, vt, i_{l}+1}-\tilde{e}_{ct, vt, i_{l}})
+ 
\tilde{e}_{ct, vt, i_{l}}
], 
\label{technical-emission-eq-L} 
\end{equation}
where $\tilde{w}_{ct, vt, i_{l}}$  
are experimental velocities,
$|{\bf{w}}_{l, vt}^{s}(x, t)|
\in (\tilde{w}_{ct, vt, i_{l}}, \tilde{w}_{ct, vt, i_{l}+1}),$
$\tilde{e}_{ct, vt, i_{l}},$
are experimental 
emissions of the $ct$th exhaust gas from single vehicle 
of $vt$th type at velocities $\tilde{w}_{ct, vt, i_{l}}$,
respectively,  
measured in [$\frac{{\rm kg}}{{\rm veh \cdot s}}$],
$i_{l}=1, ..., N_{EM},$
$N_{EM}$ is the number of experimental measurements.
Similarly, the dependence of the change 
of the linear density of energy on the density and 
velocity of vehicles is taken in the form:

{\bf E8b.}
\begin{equation}
\sigma_{l, vt}^{s}(x, t)=
q_{vt} \cdot k_{l, vt}^{s}(x, t) \cdot
[
(\frac{|{\bf{w}}_{l, vt}^{s}(x, t)|-\bar{w}_{vt, i_{l}}}
{\bar{w}_{vt, i_{l}+1}-\bar{w}_{vt, i_{l}}}) \cdot 
(\sigma_{vt, i_{l}+1}-\sigma_{vt, i_{l}})
+ 
\sigma_{vt, i_{l}}
],
\label{technical-heat-eq-L} 
\end{equation} 
where 
$\sigma_{vt, i_{l}},$
are experimental values of consumption of gasoline/diesel for a single 
vehicle of $vt$th type at velocities $\bar{w}_{vt, i_{l}},$
respectively, 
measured in [$\frac{{\rm kg}}{{\rm veh \cdot s}}$],
$q_{vt}$ is the emitted combustion 
energy per unit mass of gasoline/diesel
[$\frac{{\rm J}}{{\rm kg}}$] (compare \cite{CORINAIR 1993}).

\section{\bf Optimization problems.}
\label{sect-optimization}
\setcounter{equation}{0}  
Our control task is the minimization of the measures 
of the total travel time (TTT) 
\cite{Michalopoulos 1984},
emissions (E), and concentrations (C) 
of exhaust gases in the street canyon, 
therefore the appropriate optimization problems 
may be formulated as follows \cite{Duras 1998 thesis}:

{\bf F0. Vector of control.}
\begin{equation}
{\bf {u}}=(g_{1}, C_{1}, g_{2}, C_{2}, F) \in U^{{\rm adm}},
\label{control-5-tuple-eq}
\end{equation}
where ${\bf {u}}$ is vector of boundary control, $g_{m}$ are green times, 
$C_{m}$ are cycle times,
$F$ is offset time, 
and $U^{{\rm adm}}$ is a set of admissible control variables
(compare {\bf A8, A9, B5, B5S, B5SS}). 

We define six functionals {\bf F1-F6} of the total travel time,
emissions, and concentrations of pollutants
in single canyon, and in canyon with the nearest neighbour substitute canyons,
respectively.

{\bf F1. Total travel time for a single canyon.} 
\begin{equation}
J_{{\rm TTT}}({\bf{u}}) 
=\sum_{s=1}^{2} \sum_{l=1}^{n_{s}} \sum_{vt=1}^{VT}
\int_{0}^{a} \int_{0}^{T_{S}}
k_{l, vt}^{s}(x, t) dx \, dt .
\label{functional-TTT-def} 
\end{equation}

{\bf F2. Global emission for a single canyon.}
\begin{equation}
J_{{\rm E}}({\bf{u}})
=\sum_{s=1}^{2} \sum_{l=1}^{n_{s}} \sum_{ct=1}^{CT} \sum_{vt=1}^{VT}
\int_{0}^{a} \int_{0}^{T_{S}}
e_{l, ct, vt}^{s}(x, t) dx \, dt .
\label{functional-E-def} 
\end{equation}

{\bf F3. Global pollutants concentration for a single canyon.}
\begin{equation}
J_{{\rm C}}({\bf{u}})=
\rho_{{\rm STP}} \cdot 
\sum_{i=1}^{N_{E}-1} 
\int_{0}^{a} \int_{0}^{b} \int_{0}^{c} \int_{0}^{T_{S}}
c_{i}(x, y, z, t) dx \, dy \, dz \, dt.
\label{functional-C-def}
\end{equation}

{\bf F4. Total travel time for the canyon in street subnetwork.}
\begin{eqnarray}
& & J_{{\rm TTT, ext}}({\bf{u}})=
J_{{\rm TTT}}({\bf{u}}) + 
\label{functional-TTT-ext-def} \\
& & +
a \cdot \sum_{s=1}^{2} \alpha_{{\rm TTT, ext}}^{s}
\sum_{l=1}^{n_{s}} \sum_{vt=1}^{VT} k_{l, vt, {\rm jam}}^{s}
\cdot (C_{s}-g_{s}).
\nonumber
\end{eqnarray}

{\bf F5. Global emission for the canyon in street subnetwork.}
\begin{eqnarray}
& & J_{{\rm E, ext}}({\bf{u}})=
J_{{\rm E}}({\bf{u}}) +
\label{functional-E-ext-def} \\
& & +
a \cdot \sum_{s=1}^{2} \alpha_{{\rm E, ext}}^{s}
\sum_{l=1}^{n_{s}} \sum_{ct=1}^{CT} \sum_{vt=1}^{VT} 
e_{l, ct, vt, {\rm jam}}^{s}
\cdot (C_{s}-g_{s}).
\nonumber 
\end{eqnarray}

{\bf F6. Global pollutants concentration for the canyon in street subnetwork.}
\begin{eqnarray}
& & J_{{\rm C, ext}}({\bf{u}})=
J_{{\rm C}}({\bf{u}}) +
\label{functional-C-ext-def} \\
& & +
\rho_{{\rm STP}} \cdot a \cdot b \cdot c \cdot 
\sum_{i=1}^{N_{E}-1} c_{i, {\rm STP}} \cdot 
\sum_{s=1}^{2} \alpha_{{\rm C, ext}}^{s} \cdot (C_{s}-g_{s}).
\nonumber
\end{eqnarray}

The integrands $k_{l, vt}^{s}, 
e_{l, ct, vt}^{s}, c_{i}$ 
in functionals {\bf F1-F6} depend on the control vector ${\bf u}$ {\bf F0}
through the boundary conditions {\bf B0-B8},
through the equations of dynamics {\bf E1-E8},
as well as, through the sources {\bf D0-D2}.
The value of the vector of control $u$ directly
affects the boundary conditions {\bf B5, B5S, B5SS},
and then the boundary conditions {\bf B6-B8}
for vehicular densities, velocities, and emissivities.
It also affects the sources {\bf D0-D2}.
Next, it propagates to the equations of dynamics {\bf E1-E8}
and then it influences the values of functionals {\bf F1-F6}.
We only deal with six monocriterial optimization problems {\bf O1-O6},
and not with one multicriterial problem.
We put the scaling parameters equal to unity:
$\alpha_{{\rm TTT, ext}}^{s}=
\alpha_{{\rm E, ext}}^{s}=
\alpha_{{\rm C, ext}}^{s}=1.0,$  
in functionals {\bf F4-F6}.
$\rho_{{\rm STP}}$ is the density of air 
at standard temperature and pressure STP,
$c_{i, {\rm STP}}$ is concentration of the $i$th
constituent of air at standard temperature and pressure.
$J_{{\rm TTT}}$ and $J_{{\rm TTT, ext}}$ are measured in [$veh \cdot s$],
$J_{{\rm E}}$ and $J_{{\rm E, ext}}$ are measured in [kg],
and $J_{{\rm C}}$ and $J_{{\rm C, ext}}$ are measured in [$kg \cdot s$],
respectively. 

Now we formulate six separate monocriterial optimization
problems {\bf O1-O6} that consist in minimization
of functionals {\bf F1-F6} with respect to control vector
{\bf F0} over admissible domain, while the equations
of dynamics {\bf E1-E8} are fulfilled.  

{\bf O1. Minimization of total travel time for a single canyon.} 
\begin{equation}
J_{{\rm TTT}}^{\star}=J_{{\rm TTT}}({\bf{u}}_{{\rm TTT}}^{\star})=\min \{ {\bf{u}}\in U^{{\rm adm}}: 
J_{{\rm TTT}}({\bf{u}}) \};
\label{optimum-TTT-def}
\end{equation}

{\bf O2. Minimization of global emission for a single canyon.}
\begin{equation}
J_{{\rm E}}^{\star}=J_{{\rm E}}({\bf{u}}_{{\rm E}}^{\star})=\min \{ {\bf{u}}\in U^{{\rm adm}}: 
J_{{\rm E}}({\bf{u}}) \};
\label{optimum-E-def}
\end{equation}   

{\bf O3. Minimization of global pollutants concentration for a single canyon.}
\begin{equation}
J_{{\rm C}}^{\star}=J_{{\rm C}}({\bf{u}}_{{\rm C}}^{\star})=\min \{ {\bf{u}}\in U^{{\rm adm}}:
J_{{\rm C}}({\bf{u}}) \};
\label{optimum-C-def}
\end{equation}   

{\bf O4. Minimization of total travel time for a canyon in street subnetwork.}
\begin{equation}
J_{{\rm TTT, ext}}^{\star}=J_{{\rm TTT, ext}}({\bf{u}}_{{\rm TTT, ext}}^{\star})
=\min \{ {\bf{u}}\in U^{{\rm adm}}: J_{{\rm TTT, ext}}({\bf{u}}) \};
\label{optimum-TTT-ext-def}
\end{equation}   

{\bf O5. Minimization of global emission for a canyon in street subnetwork.}
\begin{equation}
J_{{\rm E, ext}}^{\star}=J_{{\rm E, ext}}({\bf{u}}_{{\rm E, ext}}^{\star})
=\min \{ {\bf{u}}\in U^{{\rm adm}}: J_{{\rm E, ext}}({\bf{u}}) \};
\label{optimum-E-ext-def}
\end{equation}   

{\bf O6. Minimization of global pollutants concentration for a canyon in street subnetwork.}
\begin{equation}
J_{{\rm C, ext}}^{\star}=J_{{\rm C, ext}}({\bf{u}}_{{\rm C, ext}}^{\star})
=\min \{ {\bf{u}}\in U^{{\rm adm}}: J_{{\rm C, ext}}({\bf{u}}) \},
\label{optimum-C-ext-def}
\end{equation}   
where $J_{{\rm TTT}}^{\star}, J_{{\rm E}}^{\star}, J_{{\rm C}}^{\star}, 
J_{{\rm TTT, ext}}^{\star}, J_{{\rm E, ext}}^{\star}, J_{{\rm C, ext}}^{\star}$,
are the minimal values of the functionals {\bf F1-F6},
and ${\bf{u}}_{{\rm TTT}}^{\star}, {\bf{u}}_{{\rm E}}^{\star}, {\bf{u}}_{{\rm C}}^{\star}, 
{\bf{u}}_{{\rm TTT, ext}}^{\star}, {\bf{u}}_{{\rm E, ext}}^{\star}, {\bf{u}}_{{\rm C, ext}}^{\star}$,
are control vectors at which the functionals reach the minima, respectively.

\section{Acknowledgements}
\label{sect-acknowledgements}
\setcounter{equation}{0}  
It is my pleasure to thank Professor Andrzej Adamski
for formulating the problem,
Professors Stanis{\l}aw Bia{\l}as, Wojciech Mitkowski,
and Gwidon Szefer, for many constructive criticisms. 
I also thank Professor W{\l}odzimierz W\'ojcik
for his giving me the access to his computer facilities.

\end{document}